\documentclass[twocolumn,secnumarabic,amssymb, nobibnotes, aps, prd]{revtex4}
\usepackage{graphicx}

\begin{document}
\title{The Meissner effect puzzle and the quantum force in superconductor}
\author{A. V. Nikulov}
\affiliation{Institute of Microelectronics Technology and High Purity Materials, Russian Academy of Sciences, 142432 Chernogolovka, Moscow District, RUSSIA} 

\begin{abstract} The puzzle of the acceleration of the mobile charge carriers and the ions in the superconductor in direction opposite to the electromagnetic force revealed formerly in the Meissner effect is considered in the case of the transition of a narrow ring from normal to superconducting state. It is elucidated that the azimuthal quantum force was deduced eleven years ago from the experimental evidence of this acceleration but it can not solve this puzzle. This quantum force explains other paradoxical phenomena connected with reiterated switching of the ring between normal and superconducting states.
 \end{abstract}

\maketitle

\narrowtext

\section{Introduction}
The Meissner effect, interpreted fairly as the first experimental evidence of macroscopic quantum phenomenon, was discovered as far back as 1933. This effect has revealed a fundamental distinction between an ideal conductor and superconductor. Magnetic field should be invariable in the inside of a bulk ideal conductor in accordance with the Lenz's law whereas it changes at the Meissner effect in defiance of the Lenz's law \cite{Hirsch2007}. J.E. Hirsch expresses fairly astonishment \cite{Hirsch2010} that this puzzle is ignored: {\it "Strangely, the question of what is the 'force' propelling the mobile charge carriers and the ions in the superconductor to move in direction opposite to the electromagnetic force in the Meissner effect was essentially never raised nor answered to my knowledge, except for the following instances: \cite{London1935}  (H. London states: "The generation of current in the part which becomes supraconductive takes place without any assistance of an electrical field and is only due to forces which come from the decrease of the free energy caused by the phase transformation," but does not discuss the nature of these forces), \cite{PRB2001} (A.V. Nikulov introduces a "quantum force" to explain the Little-Parks effect in superconductors and proposes that it also explains the Meissner effect)"}. The azimuthal quantum force, which J.E. Hirsch mentioned, was introduced \cite{PRB2001} in order to explain some paradoxical phenomena observed in superconducting rings but it can not explain the Meissner effect. Its essence and the region of application will be clarified in the present paper. The puzzle of the acceleration of the mobile charge carriers and the ions against the electromagnetic force indicated by J.E. Hirsch \cite{Hirsch2010} in the Meissner effect is revealed also at the transition of a thin ring from normal to superconducting state. This and other puzzles will be considered in the next section.  The deduction of the azimuthal quantum force from the Ginzburg-Landau theory and experimental facts will be described also in this section 2. The explanation of some paradoxical phenomena with help of the azimuthal quantum force will be considered in the section 3.

\section{Deduction of the azimuthal quantum force}
The Ginzburg-Landau theory \cite{GL1950} describes superconducting state with help of the GL wave function $\Psi _{GL} = |\Psi _{GL}|\exp i\varphi $, in which $|\Psi _{GL}|^{2} = n_{s}$ is interpreted as the density of superconducting pairs and $\hbar \bigtriangledown \varphi = p = mv + qA$ is momentum of single pair with the mass $m$ and the charge $q = 2e$. Superconducting current density is 
$$j_{s} = \frac{n_{s}q}{m}(\bigtriangledown \varphi - qA) = n_{s}q v \eqno{(1)}$$ 
when $\bigtriangledown n_{s} = 0$ \cite{Tinkham}. The quantization can be deduced from the requirement that the complex wave function must be single-valued $\Psi _{GL} = |\Psi _{GL}|\exp i\varphi = |\Psi _{GL}|\exp i(\varphi + n2\pi)$ at any point in superconductor \cite{Tinkham}. Therefore, its phase must change by integral multiples of $2\pi $ following a complete turn along the path of integration, yielding the Bohr-Sommerfeld quantization 
$$\oint_{l}dl \nabla \varphi = \oint_{l}dl \frac{p}{\hbar}  = \oint_{l}dl \frac{mv + qA}{\hbar} = n2\pi \eqno{(2)}$$ 
According to the relations (1), (2) and  $\oint_{l}dl A = \Phi $ the integral of the current density along any closed path inside superconductor 
$$\mu _{0}\oint_{l}dl \lambda _{L}^{2} j_{s}  + \Phi = n\Phi_{0}  \eqno{(3)}$$  
must be connected with the integral quantum number $n$ and the magnetic flux $\Phi $ inside the closed path $l$. $\lambda _{L} = (m/\mu _{0}q^{2}n_{s})^{0.5} = \lambda _{L}(0)(1 - T/T_{c})^{-1/2}$  is the London penetration depth; $\Phi _{0} = 2\pi \hbar /q$ is the flux quantum.

\subsection{Quantization in superconducting ring}
Consider a thin ($w \ll \lambda _{L}$) cylinder or a ring with a radius $r$, section $s = wh > 0$ and pair density $n _{s} > 0$  along the whole circumference $l$. According to (3), the persistent current 
$$I_{p} = sj_{s} = \frac{q\hbar}{mr\overline{(s n _{s})^{-1}}} (n - \frac{\Phi }{\Phi _{0}}) = I_{p,A}2 (n - \frac{\Phi }{\Phi _{0}})  \eqno{(4)}$$  
must flow along the circumference $l$, when magnetic flux inside the cylinder or the ring $\Phi = BS = B\pi r^{2}$ is not divisible the flux quantum $\Phi \neq n\Phi _{0}$. Here $h$ and $w$ are the height and width of ring or cylinder wall; $\overline{(s n _{s})^{-1}} = l^{-1}\oint _{l}dl (s n _{s})^{-1}$ is the value determining the amplitude $I _{p,A} = q \hbar/2mr\overline{(s n _{s})^{-1}} $  of the persistent current in a ring with section $s$ and density $ n _{s}$ which may vary along the ring circumference $l$ \cite{PRB2001}.  

The discreteness of permitted states spectrum of superconducting ring (4) at $1/\overline{(s n _{s})^{-1}} > 0$ is beyond any doubt because of numerous experimental evidences, in particular because of numerous observations of quantum periodicity of different parameters in magnetic field $B$. The quantum periodicity is observed because of the change with magnetic flux value $\Phi $ of the integer quantum number $n$ corresponding to minimal kinetic energy 
$$E_{n} = \oint _{l} dl sn _{s} \frac{mv _{n}^{2}}{2} =  \frac{I_{p}}{q}\oint _{l} dl \frac{mv_{n}}{2} = I _{p,A}\Phi _{0} (n - \frac{\Phi }{\Phi _{0}})^{2}  \eqno{(5)}$$
of superconducting pairs. This phenomenon was observed first as far back as 1962 by W. A. Little and R. D. Parks \cite{LP1962} at measurements of the resistance of thin cylinder in the temperature region corresponding to its superconducting resistive transition. Later on, the quantum oscillations of the ring resistance $\Delta R \propto  I_{p}^{2}$ \cite{Letter07,toKulik2010}, its magnetic susceptibility $\Delta \Phi _{Ip} = LI_{p}$ \cite{PCScien07}, the critical current $I_{c}(\Phi /\Phi_{0}) = I_{c0} - 2|I_{p}(\Phi /\Phi_{0})| $ \cite{JETP07J} and the dc voltage $V_{dc}(\Phi /\Phi_{0}) \propto I_{p}(\Phi /\Phi_{0})$ measured on segments of asymmetric rings \cite{Letter07,toKulik2010,PerMob2001,Letter2003,PCJETP07,PLA2012} were observed. 

\subsection{Transition between discrete and continuous spectrum of superconducting ring states}
 The spectrum of the permitted states (5) is strongly discrete $ \Delta E_{n,n+1} = |E_{n+1} - E_{n}| \gg k _{B}T$ thanks to the enormous number $ N _{s} = \oint _{l} dl sn _{s}$ of pairs in a real superconducting ring at $T \leq T_{c}$. The energy difference between permitted states $ \Delta E_{n,n+1} \approx  I _{p,A} \Phi _{0} \approx  4 \ 10^{-21} \ J$ (5) at the real amplitude $ I _{p,A} \approx  2 \ \mu A$ of the persistent current of a real ring measured even near superconducting transition $T \approx  0.99T _{c} \approx  1.24 \ K$ \cite{JETP07J} corresponds to the value $ \Delta E_{n,n+1} / k _{B} \approx  300 \ K $ exceeding strongly the temperature of measurements $T \approx 1.24 \ K$. According to (4) and (5) the persistent current and the spectrum discreteness diminish $ \Delta E_{n,n+1} \approx  I _{p,A} \Phi _{0} = \Phi _{0} q \hbar/mr\overline{(s n _{s})^{-1}}  \rightarrow 0$ with pair density decrease $1/\overline{(s n _{s})^{-1}}  \rightarrow 0$. Both the persistent current $I _{p,A} = q \hbar/2mr\overline{(s n _{s})^{-1}} $ and the energy difference $ \Delta E_{n,n+1} $ equal zero when pair density is zero $n _{s,A} = 0$ at least in a ring segment $l _{A}$, Fig.1, because $1/\overline{(s n _{s})^{-1}} \approx   s ln _{s,A} n _{s,0}/( l n _{s,A} + l _{A}n _{s,0} - l _{A}n _{s,A}) = 0$.  

\begin{figure}[b]
\includegraphics{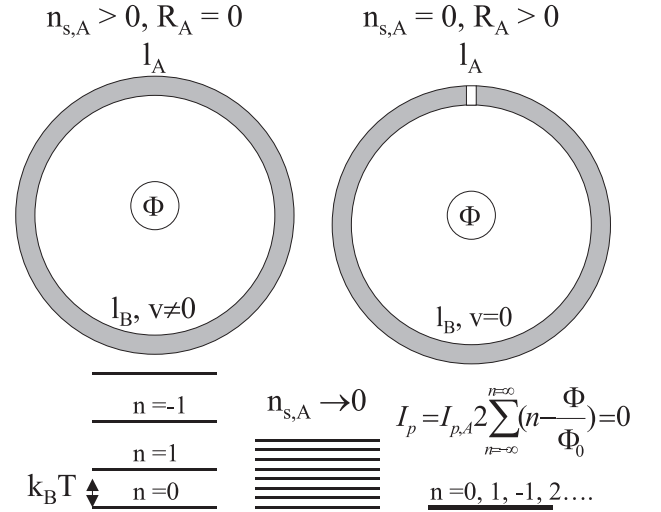}
\caption{\label{fig:epsart} The velocity of pairs and the persistent current (4) should be non-zero at $\Phi \neq n\Phi _{0}$ when the whole ring is superconducting and the spectrum of the permitted states (5) is strongly discrete $ \Delta E_{n,n+1} \gg k _{B}T$ (on the left). The spectrum of the permitted states of superconducting pairs is continuous, their velocity and the persistent current are zero at any $\Phi $ value when superconductivity is broken $n _{s,A} = 0$ in a ring segment $l _{A}$ (on the right). }
\end{figure}

\subsection{The puzzles} 
The transition between the discrete spectrum with $I _{p,A} \neq 0$ and continuous spectrum with $I _{p,A} = 0$, deduced on the base of the Ginzburg-Landau theory, was observed in numerous experiments. For example, it is observed at measurements of the critical current $I_{c}$ \cite{JETP07J,PCJETP07,NANO2011} when the external current $I_{ext}$ used for $ I _{c} $ measuring  switches at $I_{ext} = I_{c}$ the ring in the normal state $ n _{s} = 0$ at each measurement and the ring comes back in superconducting state $ n _{s} \neq 0$ when the $I_{ext}$ value decreases down to zero, see Fig.2,3 in \cite{PCJETP07}. The quantum periodicity $I_{c}(\Phi /\Phi_{0})  $ \cite{JETP07J,PCJETP07} testifies to non-zero persistent current $I _{p,A} \neq 0$ in superconducting state whereas in the normal state the electrical current $I(t) = I_{p}\exp -t/\tau_{RL}$ circulating in a ring with an inductance $L$ should decay during the relaxation time $\tau_{RL} = L/R$ because of a non-zero resistance $R > 0$. The time of each measurement of  the critical current $I_{c}$ \cite{JETP07J,PCJETP07,NANO2011} is much longer than the relaxation time $\tau_{RL} = L/R$. Therefore the current changes from $I = 0$ to $I = I _{p} = I _{p,A}2(n - \Phi /\Phi_{0})$ at each return of the ring in superconducting state. 

The mobile charge carriers accelerate in direction opposite to the electromagnetic force in this case as well as in the Meissner effect. Their angular momentum $M _{p} = (2m/q)I S $ changes from $M _{p} = 0$ to $M _{p} = (2m/q)I _{p}S = (2m/q)I _{p,A}\pi r^{2}2(n - \Phi /\Phi _{0}) $. It is well known  that the electrical current should decay at $R > 0$ because of the dissipation force $F _{dis}$ acting between electrons and the crystalline lattice of ions. Therefore the angular momentum of both the mobile charge carriers and the crystalline lattice of ions change after the transition of the ring in the normal state. We should think that the both one should change also at the transition in superconducting state, although no force acting between mobile charge carriers and the crystalline lattice of ions could be indicated in this case. Without this balance it could be possible to have rotated a ring merely with help of its switching between superconducting and normal states at $\Phi \neq n\Phi _{0}$ inside it. The angular momentum of superconducting pairs should change although without a known force but, at least, because of the known cause - the quantization (2,3). The angular momentum of ions should change even without this cause in order the ring could not rotate. 

Thus, the puzzle indicated by J.E. Hirsch \cite{Hirsch2010} in the Meissner effect is observed also at the ring switching from normal to superconducting state. In the second case this puzzle is even more mysterious. The direction, clockwise or anticlockwise, in which the mobile charge carriers accelerate against the electromagnetic force, depends only on the direction of magnetic field in the Meissner effect whereas in the ring this direction depends on the value of magnetic field $B$. The experiments testify that if, for example, this direction is clockwise at $BS = \Phi \approx  0.25\Phi _{0}$ than it will be anticlockwise at $BS = \Phi \approx  0.75\Phi _{0}$. A theory must be very ingenious in order to include a force the direction of which could change with a value. Other question, which can puzzle any theorist, may be connected with possibilities to close the wave function $\Psi _{GL} = |\Psi _{GL}|\exp i\varphi $ in a ring segment $l _{A}$, Fig.1. The quantum theory predicts even a mechanical force, which can be measured, at a mechanical closing of the superconducting ring, see the end of \cite{PRB2001}. The force should act because of the energy difference between superconducting states of the ring with zero and non-zero velocity of pairs \cite{PRB2001}. But no theory can answer on the questions: "Why and how quickly can the velocity of pairs in a segment $l _{B}$ change from $v = 0$ to $v \neq 0$ after the transition of the $l _{A}$ segment into the superconducting state $n _{s,A} > 0$?", Fig.1. This puzzle is more terrible than the acceleration of pairs in direction opposite to the electromagnetic force. The cause (the transition of the $l _{A}$ segment in superconducting state) and the effect (the change of pair velocity in the $l _{B}$ segment) may be spatially separated, Fig.1. Therefore a theory must provide a defined time during which the transition of the $l _{A}$ segment can change the velocity in the spatially separated segment $l _{B}$. I think it is easier to measure this time experimentally than to find a theory which could include this time.

\subsection{Momentum change in a time unite} 
I did not lay claim in \cite{PRB2001} to a solution of these puzzles, as well as of the Meissner effect puzzle. The azimuthal quantum force was introduced in \cite{PRB2001} in order to explain some other puzzles  observed when ring segments or the whole ring are switched many times $N _{sw} \gg 1$, with a frequency $f _{sw} = N _{sw}/\Theta $, between normal and superconducting state. The azimuthal quantum force does not explain why the mobile charge carriers and the ions can accelerate against the electromagnetic force. It is deduced from this puzzle. Hirsch argues \cite{Hirsch2010} {\it "that there is no physical basis for such an azimuthal force"}. I agree with this assertion if only the universally recognised quantum formalism, in particular the Ginzburg-Landau theory, and above all numerous experimental results can not be considered as the physical basis. The quantum formalism  (4) states, in full accordance with all experimental results, that the persistent current must appear at the closing of superconducting state in the ring, Fig.1. The change $\hbar (n -\Phi /\Phi _{0})$ of the angular momentum of each pair (at $v = 0$ before the closing) must be because of the quantization demand $rp = n \hbar $ (2). Although the quantum formalism can not explain how this change of macroscopic number $N_ {s} = 2 \pi r sn _{s}$ of pairs can be possible without any force, all experimental results testify to this force-free angular momentum transfer.

Consider the switching with a frequency $f _{sw}  \ll 1/\tau_{RL}$ of the whole ring between superconducting and normal states observed, for example, at the measurements of the critical current $I_{c}$ \cite{JETP07J,PCJETP07,NANO2011}, see the section 2.3. In the superconducting state the circular electrical current equals $I = I _{p} = I _{p,A}2(n - \Phi /\Phi_{0})$ and the velocity of pairs $ \oint_{l}dl v = 2\pi r v = (2\pi \hbar /m) (n - \Phi /\Phi _{0})$ because of the quantization (2,4) whereas in normal state $I = 0$ and the average velocity $v = 0$. Consequently the angular momentum of mobile charge carriers changes on $\Delta M _{p} = \pm M _{p} = \pm (2m/q)I _{p}S = \pm (2m/q)I _{p,A}\pi r^{2}2(n - \Phi /\Phi _{0}) $ at each switching in normal ($\pm = -$) or superconducting ($\pm = +$) state. The angular momentum of each electron pair $m_{p} = mvr$ changes on $\Delta m_{p} = \pm \hbar (n - \Phi /\Phi _{0})$. The average velocity of mobile charge carriers falls down to zero $\oint dl v = (2\pi \hbar /m) (n - \Phi /\Phi _{0})\exp {-t /\tau_{RL}} $ because of the dissipation force $F _{dis} = -\eta v $ in accordance with the Newton's second law $mdv/dt = F _{dis}$. Here the relaxation time may be written $\tau_{RL} = m/\eta $. The angular momentum of electron pair varies from $mvr = m\oint dl v/2\pi = \hbar (n - \Phi /\Phi _{0})$ to $mvr = 0 $ during a time $t \gg \tau_{RL}$ in accordance with the relation $\int _{0}^{t} dt \oint dl F _{dis}/2\pi = -\eta \int _{0}^{t}dt \oint dl v/2\pi = -(\eta /m) \hbar (n - \Phi /\Phi _{0}) \int _{0}^{t}dt \exp {-t /\tau_{RL}} \approx  -\hbar (n - \Phi /\Phi _{0}) $ after each transition into normal state. Consequently the dissipation force on average in time $\overline{F _{dis}} = \Theta ^{-1}\int _{0}^{\Theta }dt F _{dis}$ may be described with the relation $\oint _{l}dl \overline{ F _{dis}}/2\pi =  -\hbar (\overline{n} -\Phi /\Phi _{0}) f _{sw} $ when the ring is switched with a frequency $1/\Theta \ll f _{sw}  \ll 1/\tau_{RL}$ between superconducting and normal states. The experiments \cite{JETP07J,PCJETP07,NANO2011} testify to a non-zero average value $\overline{ F _{dis}}$ at $\Phi \neq n\Phi _{0}$ and $\Phi \neq (n+0.5)\Phi _{0}$ because of the predominant probability $P _{n} \propto \exp{- E_{n}/ k _{B}T} $ of the superconducting state with the same number $n$ at $(n-0.5)\Phi _{0} < \Phi < (n+0.5)\Phi _{0} $.

The dissipation force $\oint _{l}dl \overline{ F _{dis}}/2\pi =  -\hbar (\overline{n} -\Phi /\Phi _{0}) f _{sw}$, does not change the angular momentum of mobile charge carriers during a long time $ \Theta \gg 1/f_{sw}$ because the angular momentum must revert to the permitted value (2) at each return to superconducting state. This angular momentum change in a time unity 
$$\hbar (\overline{n} -\frac{\Phi }{\Phi _{0}}) f _{sw} = rF _{q} \eqno{(6)}$$  
compensates 
$$\oint _{l}dl \overline{ F _{dis}} + 2\pi rF _{q}=  0 \eqno{(7)}$$
the one because of the dissipation force. The change of the pair momentum in a time unity $F _{q}$ because of the quantization (2) was called in \cite{PRB2001} quantum force. Thus, the deduction of the quantum force in \cite{PRB2001} do not overstep the limits of the universally recognised quantum formalism  and is based on experimental results.

Hirsch states that {\it "an azimuthal quantum force acting on electrons only would change the total angular momentum of the system, violating the physical principle of angular momentum conservation"} \cite{Hirsch2010}. It is not quite correct to say that the quantum force introduced in \cite{PRB2001} acts. This "force" (6) describes only momentum change in a time unity because of the quantization (2) observed at switching of a ring from normal to superconducting state or between superconducting states with different connectivity of the wave function, Fig.1. The angular momentum change $\Delta M _{p} = (2m/q)I _{p}S $ of mobile charge carriers is observed at the transition into superconducting state with $I _{p} = I _{p,A}2(n - \Phi /\Phi_{0}) \neq 0$, for example in \cite{JETP07J,PCJETP07,NANO2011}. The angular momentum change of the crystalline lattice of ions did not observed directly for the present. But we should think that the angular momentum of both mobile charge carriers and the crystalline lattice of ions should change at the transition into superconducting state as well as at the transition into normal state. Therefore if it could be said that the quantum force acts then one should assume that it, as well as the dissipation force, acts both on mobile charge carriers and the crystalline lattice of ions. 

\section{Phenomena described with help of the azimuthal quantum force}
The azimuthal quantum force was introduced in \cite{PRB2001} in order to explain why the persistent current can not decay in spite of non-zero resistance. This paradoxical phenomenon was observed first in the Little-Parks experiment \cite{LP1962}. The observations of the quantum oscillations of the resistance, $\Delta R \propto  \overline{I_{p}^{2}}$ \cite{LP1962,Letter07,toKulik2010} and magnetic susceptibility $\Delta \Phi _{Ip} = L\overline{I_{p}}$ \cite{PCScien07} give evidence that the persistent current can not decay in spite of non-zero resistance without the Faraday electrical field $-dA /dt = 0$. It is well known that an electrical current must rapidly decay $I(t) = I_{0}\exp -t/\tau_{RL}$ in a ring with a resistance $R > 0$ if magnetic flux inside the ring does not change in time $d\Phi /dt = 0$. But the persistent current does not decay!  

\subsection{Why the persistent current can not decay} 
In order to explain this paradox it was taken into account in \cite{PRB2001} that the persistent current of superconducting pairs is observed at $R > 0$ only in a narrow temperature region $T _{c} - \delta T _{c}/2 < T < T _{c} + \delta T _{c}/2 $ near superconducting transition. The resistance in this region is non-zero $R(T) > 0$ but lower than the ring resistance in the normal state $R(T) < R _{ns}$ because of the thermal fluctuations \cite{Tinkham} which switch ring segments  between superconducting and normal states. The non-zero persistent current $I_{p} \neq 0$ can be observed at $R(T) > 0$ because the values on average in time $\overline{I_{p}} = \Theta ^{-1}\int _{0}^{\Theta }dt I_{p}(t)$, $\overline{R} = \Theta ^{-1}\int _{0}^{\Theta }dt R$ are measured. The average resistance is not zero $\overline{R} > 0$ because of the switching, from time to time, of ring segments or the whole ring into the normal state and $\overline{I_{p}} \neq 0$ is observed because of the opposite switching into superconducting state with the discrete spectrum of permitted states, see Fig.1. Below the fluctuation region, in superconducting state the resistance $R = 0$ permanently and therefore the observations of the persistent current is not paradoxical whereas above this temperature region the persistent current of superconducting pairs is not observed.  
 
The electrical current decays $I(t) = I_{p}\exp -t/\tau_{RL}$ because of energy dissipation during a time $t _{n}$ when the ring resistance $R(t) > 0$. The current can diminish down to zero at $t _{n} \gg \tau_{RL}$ or a non-zero value $I(t _{n}) = I_{p}\exp - t _{n}/\tau_{RL}$ at $t _{n} \leq \tau_{RL}$ before the recurrence of the whole ring into superconducting state at $t = t _{n}$. After this recurrence the current must revert to the permitted value $I _{p} = I _{p,A}2(n - \Phi /\Phi_{0})$ (4) because of the quantization (2,3). The angular momentum of mobile charge carriers changes from $M _{p} = (2m/q)I _{p}S $ to  $M _{p} = (2m/q)I(t _{n})S $ because of energy dissipation and from $M _{p} = (2m/q)I(t _{n})S $ to $M _{p} = (2m/q)I _{p}S $ because of the quantization (2,3). Therefore the dissipation force on average in time $\overline{F _{dis}}$ should be equal the quantum force (7) although the quantum force can not be described with the relation (6) if $t _{n} \leq \tau_{RL}$, i.e. if $f _{sw}  > 1/\tau_{RL}$.   

The observations of the periodicity in magnetic field of the resistance $\Delta R \propto  \overline{I_{p}^{2}}$ \cite{LP1962,Letter07,toKulik2010} and the magnetic susceptibility $\Delta \Phi _{Ip} = L\overline{I_{p}}$ \cite{PCScien07} are evidence of  the strong discreteness $\Delta E_{n,n+1} \gg k _{B}T$ of the spectrum of superconducting state with the closed wave function, Fig.1, even in the region $ T _{c} - \delta T _{c}/2 < T < T _{c} +\delta T _{c}/2 $ where the pair density is not zero $\overline{n _{s}} > 0$ because of thermal fluctuations. These observations testify to non-zero average values of the persistent current $\overline{I_{p}} = \overline{I _{p,A}}2\overline{(n - \Phi /\Phi _{0})}  \neq 0$, the dissipation force $\overline{F _{dis}}$ and the quantum force $F _{q} \neq 0$ at $\Phi \neq n\Phi _{0}$ and $\Phi \neq (n+0.5)\Phi _{0}$. These values equal zero at $\Phi = n\Phi _{0}$ when the pairs velocity $v \propto n - \Phi /\Phi _{0} $ and the persistent current (4) equal zero in the permitted state with the minimal energy (5). The state with $v = 0$ and $ I_{p} = 0$ is forbidden at $\Phi \neq (n+0.5)\Phi _{0}$, but $\overline{I_{p}} = 0$, $\overline{F _{dis}} = 0$, $F _{q} = 0$ because the permitted states $n$ and $n+1$ have the same energy (5) $E_{n+1} = I _{p,A}\Phi _{0} (n +1 - \Phi /\Phi _{0})^{2} = I _{p,A}\Phi _{0}0.5^{2} = E_{n} = I _{p,A}\Phi _{0}(n - \Phi /\Phi _{0})^{2} = I _{p,A}\Phi _{0}(-0.5)^{2}$ (5) and the same probability $P_{n+1} = P_{n}$ but opposite directed persistent current $I_{p}(n+1) = I _{p,A}2 (n+1 - \Phi /\Phi _{0}) = I _{p,A}2(n+1 - n - 0.5) = + I _{p,A}$, $I_{p}(n) = I _{p,A}2(n - n - 0.5) = - I _{p,A}$. The observations of the oscillations of the magnetic susceptibility $\Delta \Phi _{Ip} = L \overline{I_{p}} $ \cite{PCScien07} corroborate that $\overline{I_{p}} = 0$ both at $\Phi \neq n\Phi _{0}$ and $\Phi \neq (n+0.5)\Phi _{0}$. The maximum of the resistance is observed at $\Phi \neq (n+0.5)\Phi _{0}$ \cite{LP1962,Letter07,toKulik2010} because $\Delta R \propto  \overline{I_{p}^{2}} $: $\overline{I_{p}^{2}} = \overline{I _{p,A}^{2}}4\overline{(n - \Phi /\Phi _{0})} \approx \overline{I _{p,A}^{2}}4(P_{n+1}0.5^{2} + P_{n}(-0.5)^{2}) \approx \overline{I _{p,A}^{2}} $.

A conventional electrical current $I$ can not decay at $R > 0$ only if magnetic flux inside the ring changes in time $RI = -d\Phi /dt $. The current is maintained in this case by the Faraday electrical field $E = -dA/dt$. The electrical force $F _{e} = qE $ equilibrates the dissipation force $\oint _{l}dl F _{dis} + \oint _{l}dl F _{e} =  \oint _{l}dl F _{dis} + 2\pi rq E = 0$. This relation resembles (7). One may say that the azimuthal quantum force $F _{q}$  maintains the persistent current observed at $R > 0$ and $d\Phi /dt = 0$ \cite{LP1962,Letter07,toKulik2010,PCScien07} instead of the electrical force $F _{e} = qE$.  The relation between the quantum force and of the product of the ring resistance and the current on average in time 
$$\frac{2\pi rF _{q}}{q}  = \overline{RI} \eqno{(8)}$$ 
may be obtained from (7). Although this relation resembles the Ohm's law it can not connect directly the values $\overline{I}$ and $\overline{R}$ measured, for example in \cite{Letter07,toKulik2010}, with the value of the quantum force because $\overline{RI} \neq \overline{R} \times \overline{I} $ in the general case. The dissipation force acts and $R(t)I(t) \neq 0$ only during a transitional time $t _{tr} \leq \tau_{RL}$ after the switching of ring segments into normal state whereas the ring can be also in the normal state with $R > 0$ and $I = 0$ and in superconducting state with $I _{p} \neq 0$ and $R = 0$. The approximate equality $\overline{RI} \approx  \overline{R} \times \overline{I} $  is possible only in the limit case  $f _{sw} \gg  \tau_{RL}$. Nevertheless the observations $\overline{I _{p}} \neq 0$ at $\overline{R} > 0$ \cite{LP1962,Letter07,toKulik2010,PCScien07} can not be possible at $\overline{RI} = 0$ because the transitional processes must be from  $I _{p} \neq 0$ at $R = 0$ to $|I| < |I _{p}| $ at $R > 0$.

The explanation with help of the quantum force of the observations \cite{LP1962,Letter07,toKulik2010,PCScien07} of the electrical current $\overline{I _{p}} \neq 0$ circulating in the ring without a decay at $\overline{R} > 0$ and $ d\Phi /dt = 0$ is important for the true interpretation of this paradoxical phenomenon. The dissipation force must be zero without the quantum force (7) and one should assume that the electrical current can flow without energy dissipation at non-zero resistance $R > 0$. Such assumption is preposterous. It presupposes particularly that electrical currents can have different nature. No one can doubt that the energy dissipation with the power $RI^{2}$ is observed in a ring with the resistance $R$ when the current $I$ is induced by the Faraday electrical field $RI = -d\Phi /dt $. This current can be induced in the rings in which the persistent current $\overline{I _{p}} \neq 0$ at $\overline{R} > 0$ was observed \cite{LP1962,Letter07,toKulik2010,PCScien07}. How can we discern qualitatively these electrical currents if we assume that the persistent current can be dissipationless at $\overline{R} > 0$? In spite of the absurdity of this assumption some authors \cite{PCScien09,Birge2009} interpret the persistent current observed \cite{PCScien09} in normal metal ring as dissipationless. This preposterous \cite{Birge2009} interpretation not only contradicts to the experimental results \cite{PCScien09} but also is useless \cite{toKulik2010} because of the observations of the electrical current flowing against electrical field \cite{Letter07} and the dc voltage oscillating in magnetic field just as the persistent current \cite{Letter07,toKulik2010,PLA2012}.

\subsection{Electrical current flowing against electrical field} 
The Little-Parks  oscillations \cite{LP1962} are observed \cite{Letter07,toKulik2010} with help of measurement of the dc voltage $\overline{V} = \overline{R}I _{ext}$ induced by an external current flowing from left to right (or from right to left) through the ring-halves, see Fig.1 in \cite{Letter07}. The voltage periodicity $\overline{V}(\Phi /\Phi_{0}) = \overline{R}(\Phi /\Phi_{0})I _{ext}$, see, for example, Fig.3 in \cite{Letter07}, is evidence of the persistent current $\overline{I _{p}} \neq 0$ at $\Phi \neq n\Phi_{0}$ and $\Phi \neq (n+1)\Phi_{0}$ flowing against the electrical field  on average in time $\overline{E} = - \bigtriangledown \overline{V} $ in one of the ring-halves. The oscillations $\overline{V}(\Phi /\Phi_{0}) = \overline{R}(\Phi /\Phi_{0})I _{ext}$ are observed at different values of the external current $I _{ext}$ both much higher, see Fig.3 in \cite{Letter07}, and much lower, see \cite{toKulik2010}, than the amplitude of the persistent current $I _{p,A}$. The total electrical current is $\overline{I _{n}} = 0.5I _{ext} + \overline{I _{p}}$  in one of the ring-halves and $\overline{I _{w}} = 0.5I _{ext} - \overline{I _{p}}$ in the other one, in which the $\overline{I _{p}}$ direction is opposite to the $I _{ext}$ one, see Fig.1 in \cite{Letter07}. The observations $V(\Phi /\Phi_{0}) = R(\Phi /\Phi_{0})I _{ext}$ at $I _{ext} < \overline{I _{p}}$ \cite{Letter07,toKulik2010} give evidence that the total direct electrical current $\overline{I _{w}} = 0.5I _{ext} - \overline{I _{p}}$ flows against the direct electrical field $E = - \bigtriangledown \overline{V} \approx -\overline{V}/\pi r$ in one of the ring-halves, i.e. that this ring-half is a dc power source, the power of which $W _{dc} \approx - \overline{I _{p}V}$ at $I _{ext} \ll \overline{I _{p}}$. The observations \cite{Letter07,toKulik2010} of this dc power source and the dissipation of its power $W _{dc} \approx \overline{I _{p}V}$ in the other ring-half disprove once and for all the preposterous interpretation \cite{PCScien09,Birge2009} of the persistent current $\overline{I _{p}} \neq 0$ observed at $\overline{R} > 0$ as a dissipationless phenomenon.  

In order to explain the paradoxical observations \cite{Letter07,toKulik2010} of the electrical current flowing against electrical field the switching of ring segments or the whole ring between superconducting and normal states in the fluctuation region $ T _{c} - \delta T _{c}/2 < T < T _{c} +\delta T _{c}/2 $ should be taken into account. The voltage $\overline{V} = \overline{R}I _{ext} \approx  R _{ns} t _{n}f _{sw} I _{ext}$ can be observed because of non-zero resistance during a time $t _{n}$. Mobile charge carriers can move against the electrical field $E$ during the relaxation time under its own inertia (kinetic inductance $L$) in accordance with the Newton's second law $mdv/dt = F _{dis} - qE$. It is possible over and over again thanks to the non-zero velocity corresponding to the permitted state (2) before each transition into normal state. Consider these processes when the whole ring is switched between superconducting state with a homogeneous density of pairs $n _{s} > 0$ and normal state with $n _{s} = 0$ in the whole ring. Let this symmetric ring, see Fig.1 in \cite{Letter07}, has the same sections $s _{n} = s _{w}$, the same inductance $L _{n} = L _{w} = L/2$ and the same resistance in the normal state $R _{n,ns} = R _{w,ns} = R _{l,ns}/2$ of the ring-halves. The currents 
$$I _{n} = I _{ext}/2 + I _{p}; \ \ \ I _{w} = I _{ext}/2 - I _{p} \eqno{(9)} $$ 
should flow in the ring-halves when the whole ring is superconducting, see Fig.1 in \cite{Letter07}. $I _{p} = I _{p,A}2(n - \Phi /\Phi _{0})$ is the circular persistent current (4); $ I _{ext}$ is the direct ($dI _{ext}/dt = 0$) external current flowing from left to right. Here and below the left-to-right direction corresponds to positive values of $I _{ext}$, $I _{n}$, $I _{w}$ and the clockwise direction for $I _{p}$, see Fig.1 in \cite{Letter07}. The currents in the ring-halves after the switching of the whole ring in normal state can be calculated from the relations 
$$L _{n}\frac{dI _{n}}{dt} + R _{n,ns}I _{n} = V; \ \ \ L _{w}\frac{dI _{w}}{dt} + R _{w,ns}I _{w} = V  \eqno{(10)} $$ 
The sum $L _{n}dI _{n}/dt + L _{w}dI _{w}/dt + R _{n,ns}I _{n} + R _{w,ns}I _{w} = (R _{l,ns}/2)I _{ext} = 2V$ of the relations (10) gives the voltage $V = (R _{l,ns}/4)I _{ext}$ in the normal state because the sum $I _{n} + I _{w} = I _{ext}$ can not change in time $d(I _{n} + I _{w})/dt = dI _{ext}/dt = 0$. The relation    
 $$\frac{L}{2}\frac{d(I _{n}- I _{w})}{dt} +\frac{R _{l,ns}}{2}(I _{n}- I _{w}) = 0 \eqno{(11)} $$
obtained from (10) describes the decay $I _{cir}(t) = I _{p}\exp {-t /\tau_{RL}}$ of the circular current $I _{cir} = (I _{n}- I _{w})/2$ after the switching in normal state at $t = 0$. There is taken into account that according to (9) $I _{cir} = (I _{n}- I _{w})/2 = I _{p}$ before the switching at $t = 0$.  
  
The currents in the ring-halves $I _{n} = I _{ext}/2 + I _{cir}$, $I _{w} = I _{ext}/2 - I _{cir}$ decay because of the dissipation from $I _{n} = I _{ext}/2 + I _{p}$, $I _{w} = I _{ext}/2 - I _{p}$ to $I _{n} = I _{ext}/2 + I _{cir}(t _{n})$, $I _{w} = I _{ext}/2 - I _{cir}(t _{n})$ during the time $t _{n}$ when the ring is in the normal state and revert to the initial values at the recurrence in superconducting state. Therefore $\overline{ dI _{n}/dt } = \Theta ^{-1}\int _{0}^{\Theta }dt dI _{n}/dt = (I _{n}(\Theta )- I _{n}(0))/ \Theta \approx 0$ and $\overline{ dI _{w}/dt } \approx 0$ when the ring is switched with a frequency $f _{sw} \gg 1/\Theta $ between superconducting and normal states. Taking into account that the decay of the currents because of the dissipation force $\overline{R _{n}I _{n}}$, $\overline{R _{w}I _{w}}$ is compensated with the quantization we can obtained the following relations 
$$\overline{R _{n}I _{n}} - \frac{\pi rF _{q}}{q}  = \overline{V}; \ \ \ \overline{R _{w}I _{w}} + \frac{\pi rF _{q}}{q} = \overline{V} \eqno{(12)} $$ 
connected the values on average in time. Here as well as in the relation (8) the quantum force $F _{q}$ describes the momentum change in a time unity because of the quantization at the recurrence of the ring in superconducting state. The voltage on average in time $\Theta $ can be estimate with help of the relation $\overline{V} = (R _{l,ns}/4)I _{ext} t _{n}/( t _{n}+ t _{s})$ because the resistance of the ring-halves connected in parallel $R _{n}R _{w}/( R _{n}+R _{w}) = R _{l}/4$ equals zero during a time $t _{s}$ in superconducting state and $ R _{l,ns}/4$ during a time $t _{n}$ in normal state. The $\overline{V}$ value is positive when the external current $I _{ext}$ is positive whereas $\overline{I _{w}} = I _{ext}/2 - \overline{I _{cir}} \approx I _{ext}/2 - \overline{I _{p}}$ is negative value at $I _{ext}/2 < \overline{I _{p}}$. Therefore it is obvious that the equality (12b) is not possible without the quantum force.

\subsection{Quantum oscillations of dc voltage} 
Although no theory can explain why the velocity of pairs in a segment $l _{B}$ can change from $v = 0$ to $v =  ( \hbar /mr) (n - \Phi /\Phi _{0})$ after the transition of the $l _{A}$ segment into the superconducting state $n _{s,A} > 0$, Fig.1, both quantum mechanics and all experiments testify to this paradoxical force-free momentum transfer. There can be no doubt also that the potential voltage $ V_{A}(t) = R_{A}I(t) =  R_{A}I_{p}\exp {-t /\tau_{RL}}$ should appear at each switching of the $l _{A}$ segment in the normal state with a resistance $R_{A} > 0$. Therefore, as it was shown first in \cite{JLTP1998}, potential voltage with a direct component $V _{dc}$ can be observed on the segment $l _{A}$ when it is switched between superconducting and normal states with a frequency $f _{sw} = N _{sw}/\Theta $. It is easy to calculate that this direct voltage $V _{dc} =  \int _{0}^{\Theta }dt V_{A}(t)/\Theta  $ equals approximately $V _{dc} \approx  Lf _{sw} \overline{I_{p}}$ at $f _{sw} \ll  1/\tau_{RL}$ and $V _{dc} \approx R_{A} \overline{I_{p}}$ at $f _{sw} \gg  1/\tau_{RL}$. 

The dc voltage oscillations $V _{dc}(\Phi /\Phi _{0}) \propto  \overline{I_{p}}(\Phi /\Phi _{0})$ were observed already on the ring-halves with different sections $s _{w} > s _{n}$ when the asymmetric ring is switched between superconducting and normal states by ac electrical current  \cite{Letter2003,PCJETP07} or a noise \cite{Letter07,toKulik2010,PerMob2001,PLA2012}. These paradoxical observations of the circular electrical current $\overline{I_{p}}$ flowing against the dc electrical field $E = - \bigtriangledown V _{dc}$ in one of the ring-halves can not be explained also without the quantum force. The circular electrical current $I(t)$ decays after the switching of the $l _{A}$ segment in normal state because of the dissipation $R_{A}I(t)$ inducing the electrical field $E = -\bigtriangledown V = V _{A}/( l - l _{A})$ in the other segment $l - l _{A}$. This electrical field decelerates superconducting pairs in the segment $l - l _{A}$ in accordance with the Newton's second law $mdv/dt = - qE$ and reduces the circular current from $I_{p}$ to $I(t _{n}) = I_{p}\exp {- t _{n} /\tau_{RL}}$ during the time $I(t _{n})$ when the $l _{A}$ segment is in normal state. This electrical force on average in time $\overline{qE}$ is equilibrate by the quantum force because of the recurrence to the $I_{p}$ value at each switching of the $l _{A}$ segment into superconducting state. The electrical field in the $l _{A}$ segment $E = -\bigtriangledown V = V _{A}/l _{A}$ maintains the circular current at $R_{A} > 0$ when the dissipation force acts on the mobile charge carriers. This balances of the forces on average in time give following relations 
$$\overline{RI} - \frac{ l _{A}F _{q}}{q}  = V _{dc}; \ \ \ \frac{ (l - l _{A})F _{q}}{q}  = V _{dc} \eqno{(13)}$$
explaining why the electrical current can flow against electrical field in the phenomenon of the quantum oscillations of the dc voltage $V _{dc}(\Phi /\Phi _{0}) \propto  \overline{I_{p}}(\Phi /\Phi _{0})$ observed in \cite{Letter07,toKulik2010,PLA2012}.

\section{Conclusion}
Thus, the azimuthal quantum force \cite{PRB2001} was introduced in order to compensate formally the dissipation force acting at switching of ring segments between superconducting and normal states. The taking into account of the momentum change at the switching from continuous to discrete spectrum of permitted states of the mobile charge carriers enables to explain the numerous observations of the persistent current without the preposterous and baseless claim \cite{PCScien09,Birge2009} that this electrical current can be dissipationless  at non-zero electrical resistance. The quantum force can be introduced also for the explanation of the phenomenon of the persistent current observed in normal metal rings \cite{PCScien09}. This phenomenon is observed because of the discreteness of permitted state spectrum of electrons in the ring. The dissipation force in normal metal is a consequence of electron scattering at which the spectrum of momentum permitted state becomes continuous. The persistent current does not decay in spite of the scattering and the dissipation because of the change of the average value of the angular momentum of electrons at their recurrence to the states with discrete spectrum. These changes in a time unity may be also described with help of the quantum force equilibrating the dissipation force. The quantum force is deduced from the fact of the momentum change because of the quantization (2) but it can not explain this paradoxical fact. This puzzle, which J.E. Hirsch tries to solve for the case the Meissner effect, seems more unsolvable in the cases considered here.

\end{document}